\documentclass[article,twocolumn,showpacs,superscriptaddress,10pt,pra,longbibliography]{revtex4-1} 

\usepackage{amsmath,amssymb,graphicx,bm}
\usepackage{graphicx}
\usepackage{dcolumn}
\usepackage{bm}
\begin{document}

\title{Breather solitons in highly nonlocal media}

\author{Alessandro Alberucci} \email{Corresponding author: alessandro.alberucci@gmail.com}
\affiliation{Optics Laboratory, Tampere University of Technology, FI-33101 Tampere, Finland}

\author{Chandroth P. Jisha} 
\affiliation{Centro de F\'{\i}sica do Porto, Faculdade de Ci\^encias, Universidade do Porto, R. Campo Alegre 687, Porto 4169-007, Portugal}

\author{Gaetano Assanto} 
\affiliation{N\textsl{oo}EL - Nonlinear Optics and OptoElectronics Lab, University ``Roma Tre'', Via della Vasca Navale 84, 00146 Rome - Italy}
\affiliation{Optics Laboratory, Tampere University of Technology, FI-33101 Tampere, Finland}

\begin{abstract}
We investigate the breathing of optical spatial solitons in highly nonlocal media. Generalizing the Ehrenfest theorem, we demonstrate that oscillations in beam width obey a fourth-order ordinary differential equation. Moreover, in actual highly nonlocal materials, the original accessible soliton model  by Snyder and Mitchell [Science \textbf{276}, 1538 (1997)] cannot accurately describe the dynamics of self-confined beams as  the transverse size oscillations have a period which not only depends on power but also on the initial width. Modeling the nonlinear response by a Poisson equation driven by the beam intensity we verify the theoretical results against numerical simulations.
\end{abstract}


\maketitle 

\section{Introduction}

Since the invention of the laser, optics has played an important role in nonlinear physics. One of the most known phenomena in nonlinear optics is the all-optical Kerr effect or an intensity-dependent refractive index \cite{Boyd:2009}. While in the simplest limit the change in refractive index depends on the local intensity value, in nonlocal media the nonlinear perturbation depends also on the intensity in neighboring points. Nonlocality strongly affects light propagation, leading to e.g. the stabilization of fundamental bright (2+1)D spatial solitons \cite{Dabby:1968, Suter:1993} as well as  higher-order and vector solitons \cite{Hutsebaut:2004, Alberucci:2006, Skupin:2006, Fratalocchi:2007, Buccoliero:2007, Smyth:2008_1, Buccoliero:2008,Buccoliero:2009,Izdebskaya:2015}, complex dynamics and long-range interactions of solitons \cite{Peccianti:2002, Rothschild:2006, Conti:2006} and between solitons and boundaries \cite{Alfassi:2006, Alberucci:2007_2, Peccianti:2008}. Optical nonlocality also entails the observation of fundamental phenomena, from soliton bistability \cite{Kravets:2014} to spontaneous symmetry breaking \cite{Alberucci:2015_1}, from turbulence to condensation \cite{Picozzi:2011}, from irreversibility and shock waves \cite{Barsi:07, Smyth:2008, Conti:2009, Gentilini:2015} to gravity-like effects \cite{Bekenstein:2015}. \\
In general, even in the absence of losses, self-trapped beams in nonlocal media undergoes variations in transverse size owing to a dynamic balance between self-focusing and diffractive spreading \cite{Mitchell:1999, Conti:2004}. Such behavior resembles the collective excitation phenomena  in condensed matter, e.g. the collective modes in Bose-Einstein condensates where the center of mass or the condensate size in a harmonic trap undergo oscillations \cite{Pitaevskii:2003}. In optics, if the index well associated with the nonlinear response depends only on input power, nonlinear beam propagation can be described by a linear quantum harmonic oscillator and the breathing is purely periodic \cite{Snyder:1997}. In actual media, however, self-focusing also depends on the transverse profile of the beam \cite{Conti:2004}. It was shown numerically that soliton breathing remains periodic in a (1+1)D simplified model, connecting this result with the existence of a (quasi) parabolic potential well \cite{Kaminer:2007}. \\
In this Paper we generalize the Ehrenfest theorem in order to find a set of ordinary differential equations ruling the evolution of a wave satisfying the Schr\"odinger equation. We demonstrate that, if the wave is subject to a parabolic potential, a single equation for the beam width can be derived. We then apply our new equation to the investigation of spatial optical solitons in highly nonlocal media. In such a limit, as first demonstrated by Snyder and Mitchell \cite{Snyder:1997} and later confirmed experimentally in nematic liquid crystals \cite{Conti:2004,Alberucci:2006} and thermo-optic media \cite{Rothschild:2006}, the light-induced waveguide can be satisfactorily approximated with a parabola, allowing the usage of all the mathematical tools developed for the quantum harmonic oscillator \cite{Sakurai:1994}. This important result led to the coinage of the term accessible solitons \cite{Snyder:1997}. The original model for accessible solitons predicts a breathing period depending only on the input power. Here we demonstrate that -in real media showing a non-differentiable response function- both extrema and period of the oscillations strongly depend on the input beam width. Numerical simulations with reference to a nonlinear response modeled by a Poisson equation, the latter modelling both nematic liquid crystals and thermo-optic materials \cite{Alberucci:2007}, support our findings.

\section{The Schr\"{o}dinger equation in the Heisenberg picture}

In the scalar approximation, in the harmonic regime and for small nonlinear perturbations, the paraxial propagation of an optical wavepacket $\psi$ along $z$ is governed by 
\begin{equation}    \label{eq:NLSE}
     i\frac{\partial \psi}{\partial z}+\frac{1}{2k_0 n_0} \left( \frac{\partial^2 \psi}{\partial x^2} + \frac{\partial^2 \psi}{\partial y^2} \right) +k_0 \Delta n(x,y,z) \psi =0, 
\end{equation} 
where $k_0$ is the vacuum wave-number and $n_0$ the refractive index of the unperturbed medium. Due to its formal equivalence to the Schr\"{o}dinger equation \cite{Longhi:2009}, (\ref{eq:NLSE}) states the equivalence between light propagation in space and temporal evolution of a quantum particle in a two-dimensional potential with $\hbar=1$ and an effective mass $k_0n_0$: equation (\ref{eq:NLSE}) can thus be analyzed with the tools of quantum mechanics. In the Heisenberg picture an operator $\widehat{A}$ evolves in space $z$ (or time in quantum mechanics) according to \cite{Sakurai:1994}
\begin{equation}  \label{eq:heisenberg}
\frac{d\widehat{A}}{dz}=i [\widehat{A},\widehat{H}] + \frac{\partial \widehat{A}}{\partial z},
\end{equation}
 where $\widehat{H}=\frac{\widehat{p^2_x} + \widehat{p^2_y}}{2 k_0 n_0} -k_0\widehat{\Delta n}$ is the effective Hamiltonian operator and the square brackets indicate the commutator $[\widehat{A},\widehat{B}]=\widehat{A}\widehat{B}-\widehat{B	}\widehat{A}$. In the definition of the effective Hamiltonian $\widehat{H}$, the quadratic term in the operator $\widehat{p_j}=-i\partial_{\eta}\ (\eta=x,y)$ and the term $-k_0\widehat{\Delta n}$ correspond to the effective kinetic energy and the photonic potential, respectively. Applying  equation (\ref{eq:heisenberg}) to spatial operator $\widehat{\bm{x}}$ and momentum operator $\widehat{\bm{p}}$ leads to the Ehrenfest's theorem \cite{Sakurai:1994}
\begin{equation} 
n_0 \frac{d^2\widehat{x_j}}{dz^2}= - \frac{\partial \widehat{\Delta n}}{\partial {x_j}}\   (j=1,2), 
\end{equation}
with $x_1=x$ and $x_2=y$. The Ehrenfest theorem has been successfully applied in optics to derive the trajectory of finite-size beams \cite{Jisha:2011} and the interaction of multiple filaments \cite{Conti:2006}. Here we aim to extend it and derive an ODE governing the beam width, the latter related with the operator $\widehat{A}=\widehat{x^2_j}$. Equation (\ref{eq:heisenberg}) yields 
\begin{equation} 
k_0n_0 \frac{d\widehat{x^2_j}}{dz}=\widehat{p_j}\widehat{x_j}+\widehat{x_j}\widehat{p_j}, 
\end{equation}
the latter  providing 
\begin{equation} 
\frac{k_0n_0}{2}  \frac{d^2 \widehat{x^2_j}}{dz^2}=\frac{\widehat{p^2_j}}{k_0n_0} + k_0\widehat{x_j} \frac{\partial \widehat{\Delta n}}{\partial x_j}
\end{equation}
 after a derivative with respect to $z$. Similarly, taking $\widehat{A}=\widehat{p^2}_j$ we find 
\begin{equation} 
\frac{d\widehat{p^2_j}}{dz}=k_0\left(\frac{\partial \widehat{\Delta n}}{\partial x_j} \widehat{p}_j + \widehat{p}_j \frac{\partial \widehat{\Delta n}}{\partial x_j} \right)
\end{equation}
and 
\begin{equation} 
\frac{d^2 \widehat{p^2_j}}{dz^2} = 2k_0^2 \left( \frac{\partial \widehat{\Delta n}}{\partial x_j}\right)^2 + k_0 \widehat{p}_j \frac{d}{dz}\left( \frac{\partial \widehat{\Delta n}}{\partial x_j}\right) + k_0 \frac{d}{dz}\left( \frac{\partial \widehat{\Delta n}}{\partial x_j}\right) \widehat{p}_j.
\end{equation}

The advantage of the Heisenberg picture is that the generic \textsl{bra} $ \left\langle \psi\right|$  and \textsl{ket} $\left| \psi\right\rangle $  are stationary \cite{Sakurai:1994} (invariant with $z$ in  optics), thus all the equations dealing with operators hold valid for the average values $\left\langle \widehat{A} \right\rangle_\psi= {\left\langle \psi\right| \widehat{A} \left| \psi \right\rangle}\left({\int |\psi|^2 dxdy}\right)^{-1}$, as well. For conciseness, hereafter we will omit the subscript $\psi$ when referring to average quantities related with the wave $\psi$. 

\section{Waves in a parabolic potential}

The beam trajectory obeys the Ehrenfest theorem, whereas the beam width can be obtained from two (generally vectorial) second-order ODEs in two unknowns: the width $\left\langle x_j^2\right\rangle$ of its transverse profile and the width of its Fourier transform $\left\langle p_j^2\right\rangle$. The solution is not straightforward, as  a complete knowledge of the profile $\psi(x,y,z)$ is needed to calculate the average refractive index well and its derivative. Stated otherwise, all momenta of $\psi$ - i.e., $\left\langle x^n \right\rangle$ with $n\in \mathbb{N}$ - are required to get the second momentum evolution with $z$. A substantial simplification applies when the index well is parabolic, that is,  $\Delta n(x,y,z)=\frac{a(z)}{2}(x^2+y^2)$. In this case the beam width is governed by the fourth-order ODE 

\begin{equation}  \label{eq:width_parabolic}
  \frac{n_0}{2}\frac{d^4\left\langle x_j^2 \right\rangle}{dz^4}- 2a \frac{d^2\left\langle x_j^2 \right\rangle}{dz^2} -3 \frac{da}{dz} \frac{d\left\langle x_j^2 \right\rangle }{dz} - \frac{d^2a}{dz^2} \left\langle x_j^2 \right\rangle =0.
\end{equation}

Equation (\ref{eq:width_parabolic}) must be solved with initial conditions on the beam width $\xi_0=\left\langle x_j^2 \right\rangle_0=\left\langle x_j^2 \right\rangle(z=0)$, its initial variation $\xi_1=\left.\frac{d\left\langle x_j^2 \right\rangle}{dz}\right|_{z=0}$ (vanishing in the presence of a flat phase profile), its convexity $\xi_2=\left.\frac{d^2\left\langle x_j^2 \right\rangle}{dz^2}\right|_{z=0}= \left.\frac{2\left\langle p_j^2 \right\rangle}{k_0^2 n_0^2}\right|_{z=0} + \left.\frac{2a(0) \left\langle x_j^2 \right\rangle}{n_0}\right|_{z=0}$ as well as $\xi_3=\left.\frac{d^3\left\langle x_j^3 \right\rangle}{dz^3}\right|_{z=0}= \frac{4a(0)}{n_0} \xi_1 + \frac{2a^\prime(z=0) \xi_0}{n_0}$ (prime indicates derivative with respect to $z$).
The quantity  $\xi_2$ determines the initial diffraction of the beam, with spreading depending on both the intensity profile (i.e., $\left \langle x_j^2 \right\rangle$) and the phase distribution (i.e., $\left \langle p_j^2 \right\rangle$). In free space (where $a=0$ everywhere) equation (\ref{eq:width_parabolic}) reduces to $\frac{d^4 \left\langle x_j^2 \right\rangle}{dz^4}=0$. 
Since for a real Gaussian beam of radial waist $w$ (i.e. $I(x,y)=I_0 e^{- 2(x^2 + y^2)/w^2}$) it is $\left\langle p_j^2 \right\rangle=1/w^2$, consistently with diffraction we find $w^2(z)=w^2_0 + \frac{4z^2}{k_0^2 n_0^2 w_0^2}$, where $w^2=4\left\langle x_j^2 \right\rangle$ is valid whenever the intensity profile is Gaussian.

\section{Self-trapped nonlinear waves in highly nonlocal media}

Equation (\ref{eq:width_parabolic}) is valid whenever the refractive index well is parabolic, in both linear ($a(z)$ independent of excitation) and nonlinear 	 ($a(z)$ depending on wavepacket profile and amplitude) regimes \cite{Snyder:1995}. 
In the highly nonlocal limit the light-induced index well is much wider than the beam \cite{Snyder:1997}; the photonic potential can be Taylor-expanded to the second-order and equation (\ref{eq:width_parabolic}) accurately models light propagation. We solve it in Kerr media (refractive index dependent  on  intensity $I=n_0|\psi|^2/(2Z_0)$ with $Z_0$ the vacuum impedance) with reference to two common responses: Gaussian  \cite{Snyder:1997,Krolikowski:2000} and diffusive-like \cite{Bekenstein:2015}. 
The two responses differ for the Green function $G$ linking the beam intensity $I$ to the nonlinear perturbation $\Delta n=\int\int{I(x^\prime,y^\prime)G(x-x^\prime,y-y^\prime)dx^\prime dy^\prime}$, with $\int\int Gdxdy =1$. Hereafter, for the sake of simplicity we refer to either (2+1)D structures with cylindrical symmetry or (1+1)D geometries. 

\subsection{Ideal limit: differentiable Green function}

\begin{figure}[htbp]
\centering
\includegraphics[width=0.5\textwidth]{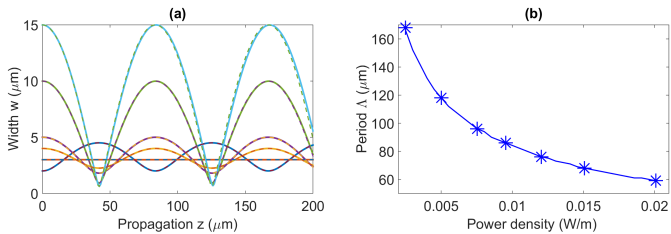}
\caption{(Color online) (a) Solitary width $w$ versus $z$ for various input widths $w_0$ (2, 3, 4, 5, 10 and 15 $\mu$m, respectively) and a power density $\mathcal{P}=0.01$ Wm$^{-1}$, corresponding to $w_S= 3~\mu$m. Numerical (solid) and analytical (dashed) lines overlap. (b) Theoretical (solid line) and numerically calculated (symbols) oscillation period $\Lambda$ versus $\mathcal{P}$ for $w_0= 3 ~\mu$m. Here the wavelength is 1064 nm.}
\label{fig:theory1D}
\end{figure}

When the Green function $G$ is twice differentiable in the origin, it is easy to obtain $\Delta n\approx \left(G_0 + 2 G_2 \left\langle x^2 \right\rangle\right) P + G_2 P\ (x^2+y^2)$, with coefficients $G_m=\frac{1}{m!}\frac{\partial^m G}{\partial x^m}$ where $G_m$ is computed in the origin. This exactly matches the Snyder-Mitchell model, with a nonlinear response exclusively dependent on input power $P$ \cite{Snyder:1997}. The term $\left(G_0 + 2 G_2 \left\langle x^2 \right\rangle\right) P$ is $x$-independent, but   it varies along $z$ through the local beam width $\propto \sqrt{\left\langle x_j^2 \right \rangle}$,  yielding an overall power-dependent  phase shift  of the beam \cite{Guo:2004}. The nonlinear lens, modeled by the transverse term proportional to $(x^2+y^2)$, is constant with $z$ because the strength  $a=2 G_2 P$ of the quantum harmonic oscillator is invariant across the sample. For planar phase fronts at the input, the beam breathing along $z$ follows \cite{Snyder:1997}
\begin{equation}  \label{eq:breathing_differentiable}
 w^2= \frac{w_0^4 + w_S^4(P) }{2w_0^2} + \frac{w_0^4 - w_S^4(P) }{2w_0^2} \cos{\left( \sqrt{\frac{8|G_2|P}{n_0}} z \right)},
\end{equation}
where $w_S(P)=\left(\frac{2}{n_0 k_0^2 |G_2| P }\right)^{1/4}$ is the radial waist of the soliton at power $P$. According to (\ref{eq:breathing_differentiable})  self-confined beams oscillate around  $w_\mathrm{av}=\sqrt{\frac{w_0^4 + w_S^4(P)}{2w_0^2}}$ and are affeced by both input power $P$ and  width $w_0$. The breathing period is proportional to $1/\sqrt{P}$, whereas the breathing amplitude increases with the difference $|w_0-w_S|$. Figure~\ref{fig:theory1D} compares the predictions of (\ref{eq:breathing_differentiable}) with BPM (Beam Propagation Method) simulations in (1+1)D (with power $P$ replaced by a power density $\mathcal{P}$ in Wm$^{-1}$), confirming that the Snyder-Mitchell model is correct in the  Gaussian limit \cite{Guo:2004}.

\subsection{Real case: singular Green function}

Since actual highly nonlocal materials obey a diffusion-like equation with a  Green function non-differentiable in the origin, 
this leads to discrepancies and quantitative inaccuracies when they are described by the original Snyder-Mitchell model \cite{Guo:2004, Alberucci:2014_1}. In thermo-optic materials or nematic liquid crystals in the perturbative regime \cite{Peccianti:2012}, for instance, the nonlinear index well in the perturbation regime stems from a Poisson equation (we neglect the $\Delta n$ derivative  along $z$ for simplicity)
\begin{equation}   \label{eq:poisson}
  \nabla_{xy}^2 \Delta n + n_2 I= 0,
\end{equation}
with the  factor $n_2$ ($n_2>0$ for self-focusing) an equivalent nonlocal Kerr coefficient accounting for the ratio between the beam amplitude and the corresponding index perturbation. From (\ref{eq:poisson}), the nonlinear index well for an arbitrary beam profile can be Taylor expanded as $\Delta n\approx \Delta n_0 - \frac{n_2 I_0}{4} (x^2 + y^2)$, with $I_0$ the intensity in the origin \cite{Conti:2004}. For a Gaussian beam it is $I_0=2P/(\pi w^2)$, thus
\begin{equation}  \label{eq:a_poisson}
  a=-\frac{ n_2 P}{\pi w^2} = - \frac{n_2 P}{4 \pi \left\langle x^2 \right\rangle}. 
\end{equation}
Substitution of (\ref{eq:a_poisson}) in equation (\ref{eq:width_parabolic}) provides 
\begin{equation}  \label{eq:second_order}
\frac{d^2}{dz^2}\left( \frac{d^2 \left\langle x^2 \right\rangle}{dz^2} + \frac{n_2 P }{2\pi n_0} \ln \left\langle x^2 \right\rangle \right)=0. 
\end{equation}
Equation (\ref{eq:second_order}) shows that the fourth-order ODE equation (\ref{eq:width_parabolic}) turns into a second-order ODE when the medium nonlinearity is governed by equation (\ref{eq:a_poisson}).
Applying the proper boundary conditions we find
\begin {equation}   \label{eq:newton}
  \frac{d^2 \left\langle x^2 \right\rangle}{dz^2} + \frac{n_2 P }{2\pi n_0} \ln \frac{\left\langle x^2 \right\rangle}{\left\langle x^2 \right\rangle_0} + \frac{2}{k_0^2 n_0^2} \left(\frac{1}{w_S^2} - \frac{1}{w_0^2} \right) = 0,
\end{equation}
with $w_S(P)=\left(\frac{4\pi}{n_0 n_2 k_0^2 P}\right)^{1/2}$. 
As expected, in the linear regime $\frac{d^2 \left\langle x^2 \right\rangle}{dz^2}=\frac{2}{k_0^2 n_0^2 w_0^2}$; when $w_0=w_S(P)$ it is $\frac{d^2 \left\langle x^2 \right\rangle}{dz^2}=0$, i.e., a shape-preserving soliton is excited. \\
Equation (\ref{eq:newton}) corresponds to the motion of a classical particle subject to a conservative force 
\begin {equation} 
F= - \frac{n_2 P }{2\pi n_0} \ln \frac{\left\langle x^2 \right\rangle}{\left\langle x^2 \right\rangle_0} - \frac{2}{k_0^2 n_0^2} \left(\frac{1}{w_S^2} - \frac{1}{w_0^2} \right),
\end{equation}
 with the latter depending  on both the normalized excitation $n_2 P$ and the initial width $w_0$. The force $F$ vanishes when $\left\langle x^2 \right\rangle= \left\langle x^2 \right\rangle_\mathrm{av}=\left\langle x^2 \right\rangle_0 \exp{\left(w^2_S/w_0^2 -1\right)}$. Then equation (\ref{eq:newton}) can be recast in the form
\begin {equation}  
\frac{d^2 \left\langle x^2 \right\rangle}{dz^2} + \frac{n_2 P }{2\pi n_0} \ln \frac{\left\langle x^2 \right\rangle}{\left\langle x^2 \right\rangle_\mathrm{av}}=0, \end{equation}
with an effective power-dependent potential acting on the beam width
\begin{equation}  \label{eq:potential}
  V(\left\langle x^2 \right\rangle,P,w_0) = \frac{n_2 P }{2\pi n_0} \left\langle x^2 \right\rangle \left(\ln \frac{\left\langle x^2 \right\rangle}{\ \ \left\langle x^2 \right\rangle_\mathrm{av}} - 1 \right).
\end{equation}

\begin{figure}[htbp]
\centering
\includegraphics[width=0.5\textwidth]{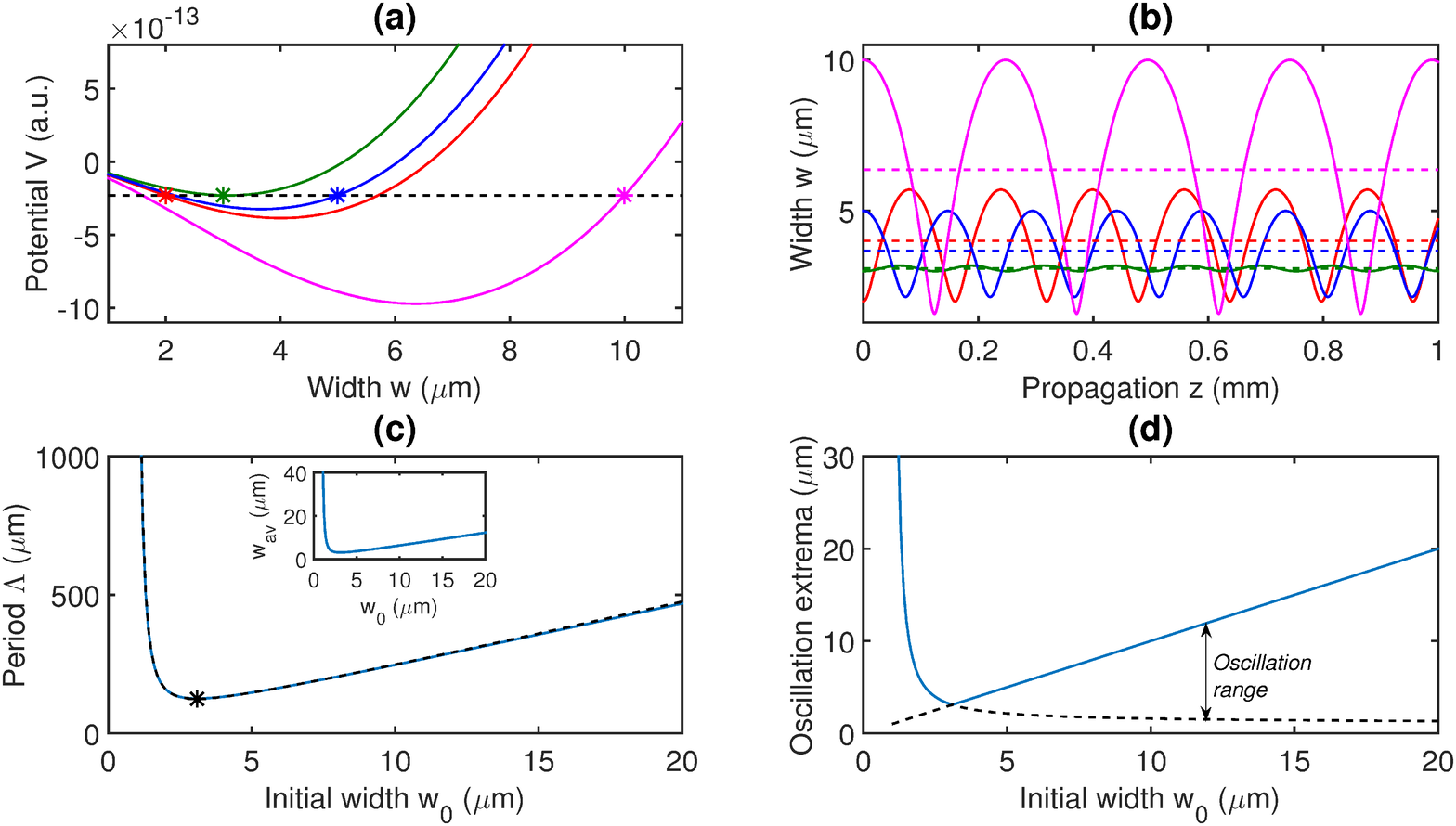}
\caption{(Color online)  (a) Potential $V$ versus beam width $w$ when the input width $w_0$ is 2 (red), 3 (green), 5 (blue) and 10$~\mu$m (magenta), respectively; stars indicate the launch conditions. (b) Width oscillation versus $z$ corresponding to the cases n (a). (c) Numerically (solid blue) and theoretically (dashed black) calculated, via equation \ref{eq:period}, oscillation period $\Lambda$ versus $w_0$; the star marks the $z$-invariant soliton. Inset: location $w_\mathrm{av}$  of the potential minimum  versus $w_0$. (d) Maximum (blue solid) and minimum (black dashed) beam width versus $w_0$. Here the wavelength is 1064nm and $n_2 P/(4\pi)\approx-0.003$, corresponding to $w_S\approx 3~\mu$m. }
\label{fig:theory}
\end{figure}


Figure \ref{fig:theory}(a) illustrates the potential $V$, asymmetric with respect to the local minimum $\left\langle x^2 \right\rangle_\mathrm{av}$ and therefore sustaining non-sinusoidal oscillations of the momentum $\left\langle x^2 \right\rangle$. Such dynamics is confirmed by direct numerical integration of equation (\ref{eq:newton}), as  plotted in figure \ref{fig:theory}(b). Integrating the energy conservation law over one half-period yields the breathing period 
\begin{equation}
\Lambda=2\int_{\left\langle x^2 \right\rangle_0}^{\left\langle x^2 \right\rangle_F}{\frac{d\left\langle x^2 \right\rangle}{\sqrt{2\left[ V({\left\langle x^2 \right\rangle_0}) - V(\left\langle x^2 \right\rangle) \right]}}}, \label{eq:period}
\end{equation}
where $\left\langle x^2 \right\rangle_F$ is the extremum opposite to the initial value during one single oscillation.
Results from (\ref{eq:period}) are graphed in figure \ref{fig:theory}(c) together with the direct numerical integration of equation (\ref{eq:newton}) (corresponding to the beam width graphed in figure \ref{fig:theory}(b)): the match is nearly perfect. At variance with equation (\ref{eq:breathing_differentiable}), the oscillation period $\Lambda$ depends not only on input power $P$ but also on input width $w_0$. In particular, $\Lambda$ has a local minimum $\Lambda_\mathrm{min}=\sqrt{\frac{4n_0\pi^3 w_S^2}{n_2 P}}=\frac{4\pi^2}{k_0} \frac{1}{n_2 P}$ when the input beam matches the shape-preserving soliton, i.e., $w_0=w_S$: small departures from $w_S$ cause the breathing to approximately follow  equation (\ref{eq:breathing_differentiable}) as the strength $a$ of the quantum harmonic oscillator (\ref{eq:a_poisson}) undergoes small variations along $z$ \cite{Conti:2004}. For input beams narrower than the exact soliton ($w_0<w_S(P)$) the period increases sharply due to the diffraction limit (beam 
size comparable with wavelength); when $w_0>w_S(P)$ the period grows linearly with $w_0$. The location $w_\mathrm{av}=2\sqrt{\left\langle x^2 \right\rangle_\mathrm{av}}$ of the minimum effective potential  follows a  trend similar to $\Lambda$ vs $w_0$ (inset of figure \ref{fig:theory}(c)), asymptotically tending to a straight line with slope $w_0/\sqrt{e}$ for $w_0 \gg w_S$. Finally, figure \ref{fig:theory}(d) plots maximum and minimum beam widths versus $w_0$: for $w_0<w_S$ the initial width $w_0$ is the minimum (diffraction overcoming self-focusing at the input); conversely, when $w_0>w_S$, $w_0$ is the maximum (self-focusing dominating over diffraction at the input).


\begin{figure}
 \centering
 \includegraphics[width=0.5\textwidth]{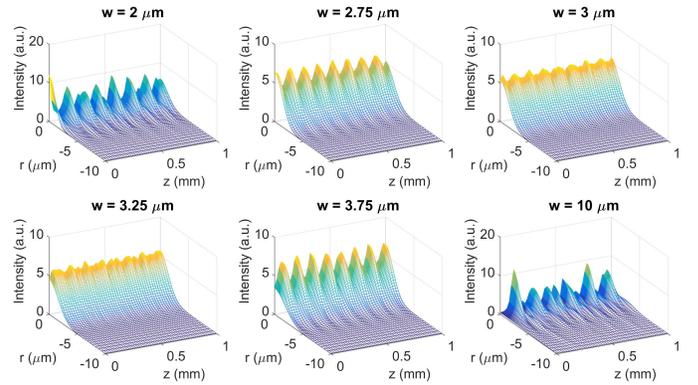}
 \caption{(Color online) Intensity profile in the plane $rz$ for various input beam widths $w_0$, the oscillations in propagation tend to zero at widths close to $w_S\approx 3~\mu$m with $n_2 P/(4\pi)\approx-0.006$. The radial extent of the sample is 100 $\mu$m and the wavelength 1064 nm.  }
 \label{fig:poisson_simulations}
\end{figure}

\begin{figure*}[htbp]
\centering
\includegraphics[width=0.95\textwidth]{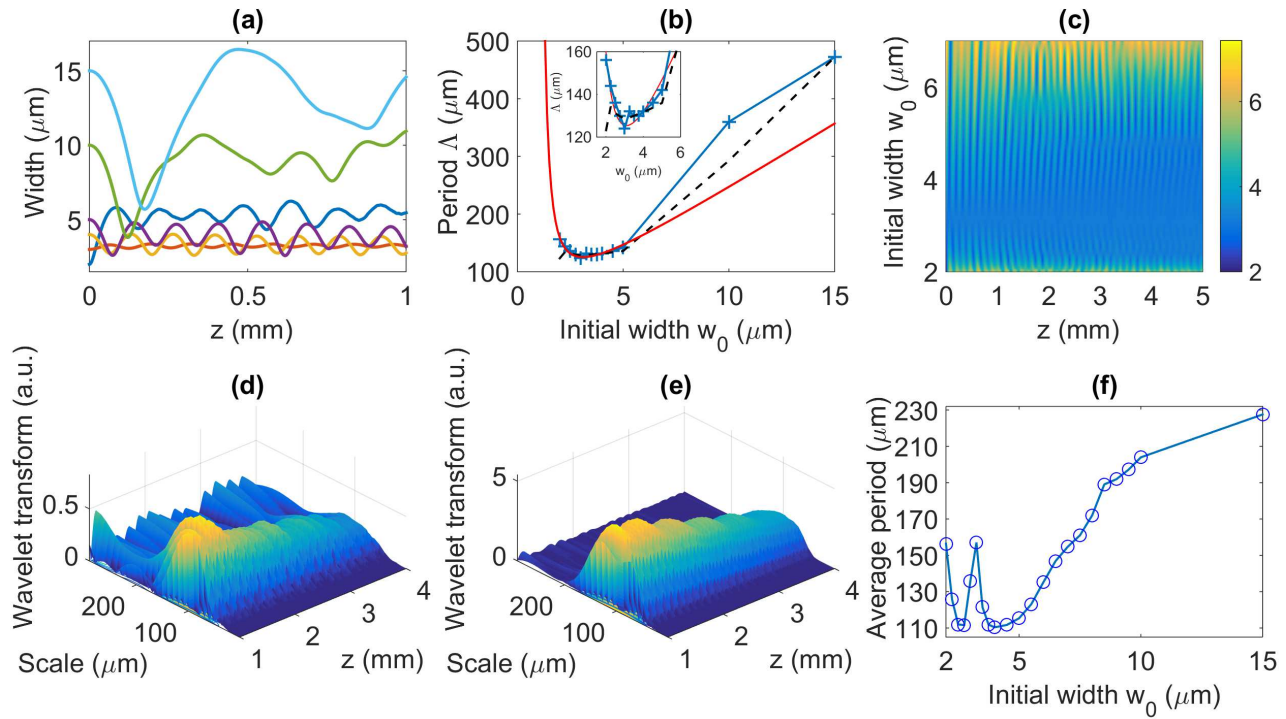}
\caption{(Color online) (a) Beam width versus $z$ when $w_0=2,\ 3,\ 4,\ 5,\ 10,$ and 15$\mu$m from $z=0$ to $z$=1 mm, respectively. (b) Oscillation period $\Lambda$ versus $w_0$: length of the first oscillation (blue line with symbols) and  average period over 1 mm (dashed line) from numerical simulations, respectively; the red solid line is the theoretical prediction from figure \ref{fig:theory}. Inset: magnification around the  minimum $w_0=3~\mu$m. (c) Color map: beam width in $\mu$m versus $z$ and initial width $w_0$; here the overall sample length is 5 mm. (d-e) Absolute value of the wavelet transform Daubechies db6 versus the scale and propagation distance $z$ when (d) $w_0=3~\mu$m and (e) $w_0=4~\mu$m. (f) Average period in the interval $1$ mm$~<z<~$5 mm computed from the wavelet transform versus $w_0$. Here $n_2 P/(4\pi)\approx-0.006$, corresponding to $w_S\approx 3~\mu$m. The radial extent of the sample is 100 $\mu$m and the wavelength 1064 nm. }
\label{fig:poisson_summary}
\end{figure*}

\subsection{Full numerical simulations in a Poisson material}

To check our predictions we integrated equations (\ref{eq:NLSE}) and (\ref{eq:poisson}) in a radially symmetric geometry, using a standard BPM in log-polar coordinates \cite{Alberucci:2014_1, Alberucci:2015}. The results for a given input power corresponding to a soliton of width $\approx 3~\mu$m are summarized in figure \ref{fig:poisson_simulations}. Noteworthy, now it is  $n_2 P/(4\pi)\approx-0.006$, i.e., the normalized input power $n_2 P$ had to be doubled with respect to the theoretical value (\ref{eq:a_poisson}) because the intensity profile overlaps with higher polynomial terms of the self-induced potential \cite{Alberucci:2014_1,Ouyang:2006}.
The intensity evolution shows a periodic to aperiodic transition for varying input widths. The case $w_0=3~\mu$m does not excite a shape-preserving soliton because in real Poisson media the exact soliton profile slightly differs from a Gaussian profile \cite{Alberucci:2014_1,Ouyang:2006}. \\
We start analyzing the wavepacket behavior when close to the input, i.e., for short propagation length. In the interval 2.5 $~\mu$m$~<w_0<5~\mu$m the excitation is close enough to the soliton state (i.e., $w_S=3~\mu$m for the chosen power) and the self-trapped beam oscillates quasi-periodically (see figure \ref{fig:poisson_simulations} and figure \ref{fig:poisson_summary}(a)). Figure \ref{fig:poisson_summary}(b) shows the first oscillation period, computed doubling the position of the first local extremum in width versus $z$. In agreement with theory the oscillation period depends on $w_0$, with $\Lambda$  shorter when $w_0\approx w_S(P)$. The numerical results resemble quite closely the predictions from equation (\ref{eq:period}), with quantitative discrepancies arising when the input beam is much wider than the soliton (see figure \ref{fig:poisson_summary}(b)). As visible in figure \ref{fig:poisson_summary}(a), on longer propagation distances the wavepacket evolution departs from theory: when the 
difference $\left| w_0-w_S \right|$ is small, the oscillation period slightly varies along $z$; conversely, both for very narrow ($w_0 \ll w_S(P)$) and very wide  ($w_0 \gg w_S(P)$) inputs, the oscillations become markedly aperiodic. The discrepancy can be ascribed to two main causes: i) the effective shape of the self-induced index well is not perfectly parabolic, as discussed above; ii) the beam shape strongly departs from Gaussian due to the nonlinear interaction between a large number of modes, in turn breaking the validity of equation (\ref{eq:a_poisson}), the relationship between $a$ and $\left\langle x^2 \right\rangle$ now requiring a more involved approach. 

The general trends with $w_0$ can be confirmed by computing light propagation over longer distances. The results in figure \ref{fig:poisson_summary}(c) show  soliton breathing over a propagation length of 5 mm. First, the evolution smoothly changes with $w_0$, ruling out the presence of chaotic dynamics \cite{Alberucci:2015,Aschieri:2013}. Second, the yellow portions in figure \ref{fig:poisson_summary}(c) (bottom and top) correspond to strongly aperiodic dynamics. Between them, in the center of the panel, the dynamics is quasi-periodic with a comb-like structure: each tooth is tilted towards the left (smaller $z$), showing that the oscillation period changes and tends to a minimum when $w_0=w_S$. Consistently with theory, the oscillation amplitude is proportional to $\left| w_0-w_S \right|$. In addition, the oscillation amplitude unexpectedly drops with $z$ due to an effective dissipation 
(in the framework of the effective potential defined via equation (\ref{eq:potential})) arising from the nonlinear interaction between the modes of the structure, as modeled by the higher-order polynomial terms in the light induced index well.  

Next we study beam breathing in the frequency domain. To carry out this analysis we use a wavelet transform, as the evolution is not periodic and extends over a finite domain. Wavelets allow to address the temporal fluctuations in the spectrum of a signal. Such goal is achieved  by using a basis composed by functions localized both in time and frequency. The family of wavelets is found by shifting and stretching a given function, named the mother wavelet. Local components of the spectrum are found compressing/dilating the mother wavelet, the compression factor used to determine at which scale we are analyzing the signal. Here we choose the wavelet transform Daubechies db6 \cite{Daubechies:1992}. To avoid artifacts due to the boundaries, we limit our analysis in the interval 1 mm$~<z<4$ mm. For both  $w_0=3~\mu$m (figure \ref{fig:poisson_summary}(d)) and  $w_0=4~\mu$m (figure \ref{fig:poisson_summary}(e)) the peak of the wavelet transform does not move on the frequency axis with $z$. The wavelet transform is 
also strongly localized on the scale axis, demonstrating that no diffusion effects occur in the frequency domain. The absolute value of the transform decreases with $z$, in line with the emerging dissipative mechanism described above. Comparing the two cases, it is evident that the spectral components are higher when $w_0=4~\mu$m due to larger oscillation amplitude. Noteworthy, for $w_0=3~\mu$m the higher frequency components  are much more relevant than for $w_0=4~\mu$m. In fact, in the former case the oscillation around the average value is heavily affected by the non-Gaussian profile of the soliton \cite{Ouyang:2006}, a contribution neglected in deriving equations (\ref{eq:newton}) and (\ref{eq:potential}). This is evident in figure \ref{fig:poisson_summary}(f) showing the average period in the range 1 mm $<z<4$ mm, computed from the wavelet transform. The shape is very close to figure~\ref{fig:poisson_summary}(b), except near $w_0=w_S$ where a spurious peak appears. Physically, close to the input the action of the higher-order modes can be neglected; for long distances their effect accumulates and cannot be neglected anymore.

\section{Conclusions}

In conclusion, using tools from quantum mechanics we derived a general equation ruling the nonlinear evolution of the beam width in a parabolic index well. Applying this model to light propagation in highly nonlocal media, we investigated how soliton breathing departs (both qualitatively and quantitatively) from the ideal Snyder-Mitchell law in real materials. In particular, we showed that the beam width dynamics can be modeled as a classic particle subject to a potential which depends on the  width of the input beam. Thus, although the beam itself introduces a longitudinal change in the index well \cite{Aschieri:2013}, remarkably no aperiodic or chaotic evolution \cite{Alberucci:2015} takes place within the validity of our model. Numerical simulations verify that the breathing period depends on the width of the input beam and confirm the absence of chaos. Moreover, the simulations indicate the emergence of novel and intriguing effects due to the nonlinear interaction of several modes, assessing the suitability of nonlocal nonlinear  optics for the investigation of many-body physics \cite{Gentilini:2015_1,Vocke:2015}.

\section*{Acknowledgments}
AA and GA thank the Academy of Finland for financial support through the FiDiPro grant no. 282858. JCP gratefully acknowledges Funda\c{c}\~{a}o para a Ci\^{e}ncia e a Tecnologia, POPH-QREN and FSE (FCT, Portugal) for the fellowship SFRH/BPD/77524/2011.


%

\end{document}